\documentclass[epj]{svjour}
\usepackage{graphics}
\usepackage[latin1]{inputenc}

\begin{document}

\title{Wüstite: Electric, thermodynamic and optical properties of FeO}

\author{F. Schrettle\thanks{e-mail: \ttfamily{florian.schrettle@physik.uni-augsburg.de}} \and Ch. Kant \and P. Lunkenheimer \and F. Mayr \and J. Deisenhofer \and A. Loidl
}                   
\institute{Experimental Physics V, Center for Electronic Correlations and Magnetism, University of Augsburg, 86135 Augsburg,
Germany}

\date{Received: date / Revised version: date}

\abstract{ We report on a systematic optical investigation of
wüstite. In addition, the sample under consideration, Fe$_{0.93}$O,
has been characterized in detail by electrical transport,
dielectric, magnetic and thermodynamic measurements. From infrared
reflectivity experiments, phonon properties, Drude-like conductivity
contributions and electronic transitions have been systematically
investigated. The phonon modes reveal a clear splitting below the
antiferromagnetic ordering temperature, similar to observations in
other transition-metal monoxides and in spinel compounds which have
been explained in terms of a spin-driven Jahn-Teller effect. The
electronic transitions can best be described assuming a
crystal-field parameter $Dq$~=~750~cm$^{-1}$ and a spin-orbit
coupling constant $\lambda$~=~95~cm$^{-1}$. A well defined crystal
field excitation at low temperatures reveals significant broadening
on increasing temperature with an overall transfer of optical weight
into dc conductivity contributions. This fact seems to indicate a
melting of the on-site excitation into a Drude behavior of
delocalized charge carriers. The optical band gap in wüstite is
close to 1.0~eV at room temperature. With decreasing temperatures
and passing the magnetic phase transition we have detected a strong blue shift of the correlation-induced band edge, which
amounts more than 15\% and has been rarely observed in
antiferromagnets.
} 

\maketitle

\section{Introduction}
\label{intro} Transition metal (TM) monoxides are prototypical
examples of strongly correlated electron systems. Their electronic
structure results from a competition between electron localization
and delocalization effects in the narrow 3$d$ bands, as formulated
in the theory of Mott \cite{Mott1949} and Hubbard
\cite{Hubbard1964}. Numerous theoretical and experimental papers
about this subject and specifically concerning TM monoxides have
been published aiming to explain the insulating properties of the
late monoxides and calculating the quasiparticle band structure and
the density of states. An exception is wüstite (FeO), which has been
far less investigated. This probably stems from the fact that
stoichiometric FeO does not exist at room temperature. It is
metastable with the tendency to decay into a two-phase mixture of
$\alpha$-Fe and magnetite Fe$_3$O$_4$. However, it is well known
that cubic Fe$_{1-\delta}$O can be synthesized at ambient pressure
for iron deficiencies ranging approximately from 0.05 $ < \delta < $
0.15:

Nonstoichiometric wüstite crystallizes in the cubic rocksalt
structure and reveals antiferromagnetic (AFM) order below the Néel
temperature $T_{\mathrm{N}} \approx$ 200~K. Early reports on
antiferromagnetism of FeO with ordering temperatures between 183 K
and 198 K have been published by Millar \cite{Millar1929}, Bizette
and Tsai \cite{Bizette1943} and Foex and Lebeau \cite{Foex1948}
utilizing thermodynamic, magnetic and thermal-expansion experiments,
respectively.  Neutron diffraction at 80 K, well below the magnetic
ordering temperature, has been performed by Shull et al.
\cite{Shull1951} These authors observed the typical AFM diffraction
pattern of the insulating transition-metal monoxides with a doubling
of the chemical unit cell. The almost complete absence of the (111)
reflection indicated that in FeO the magnetic moments of Fe$^{2+}$
are aligned perpendicular to the ferromagnetic (FM) (111) sheets
with strictly alternating moment directions. Roth confirmed this
magnetic structure and determined in addition an average iron moment
of $\mu$ = 3.32 $\mu_{\mathrm{B}}$ at 4.2 K \cite{Roth1958}. This
value is clearly reduced with respect to the expected spin-only
value for Fe$^{2+}$ ($d^6$, $S$ = 2) of about 4 $\mu_{\mathrm{B}}$,
assuming an effective $g$-value $g \approx 2$ and is difficult to
explain. Even assuming an iron deficiency $\delta$ = 0.1 still yields
a larger ordered iron moment of 3.6 $\mu_{\mathrm{B}}$. The actual
effective $g$-value for Fe$^{2+}$ in octahedral environment depends
strongly on the spin-orbit coupling, covalency, and possible
distortions of the octahedra as discussed by Goodenough, who
predicted anisotropic \textit{g}-values by considering spin-orbit
coupling and a trigonal distortion \cite{Goodenough1968b}. Such a
situation is realized in FeO, which exhibits a rhombohedral
distortion below the Néel temperature. Contrary to MnO and NiO the
angle of the unit rhomb becomes less than 60° and the cell is
elongated along the [111] axis \cite{Tombs1950} in agreement with
the direction of the antiferromagnetic axis. Upon entering the
magnetically ordered state a clear splitting of the infrared active
phonon is observed, similar to other TM monoxides
\cite{Rudolf2008,Kant2008} and a number of Cr spinels
\cite{Sushkov2005,Hemberger2006,Hemberger2007b,Rudolf2007,Kant2011}.
Theoretically, this splitting has been described in the framework of
a spin-Jahn-Teller effect \cite{Yamashita2000,Tchernyshyov2002}.

In this work we discuss in detail the optical properties of
Fe$_{0.93}$O including polar phonon modes, electronic transitions
related to the Fe$^{2+}$ ions, and the onset of the band gap.
Moreover, we find the existence of a Drude-like contribution in the
paramagnetic state and revisit the magnetic, electric, and
thermodynamic properties making use of the obtained optical
excitation spectrum.

\section{Experimental details}
\label{sec:1}

Single crystals of FeO in the form of platelets with dimensions of
approximately 10 mm$^2$ and 1 mm thickness and polished to optical
quality were purchased from MaTeck GmbH. For structural
characterization small pieces of a crushed single crystal were
investigated by x-ray powder diffraction using a STOE diffractometer
with a position sensitive detector. To measure the electrical
resistance, a standard four-point technique was utilized down to
temperatures of approximately 100 K with a constant current of 10
mA. For lower temperatures and increasingly high resistance a
constant voltage of 10 V was applied in a two-point configuration.
The latter measurements were performed with the Keithly Electrometer
6517A with an input impedance $>$ 200 T$\mathrm{\Omega}$. The dielectric
properties were measured employing a frequency-response analyzer
(Novocontrol $\alpha$-analyzer). The magnetic properties were
studied using a commercial superconducting quantum interference
device magnetometer (Quantum Design MPMS-5) with magnetic fields up
to 50 kOe and temperatures from 2 K $< T <$ 400 K. The heat capacity
was measured in a physical properties measurement system (Quantum
Design PPMS) for temperatures from 2 K $< T <$ 300 K. The
reflectivity measurements were carried out using the Bruker
Fourier-transform spectrometers IFS 113v and IFS 66v/S, both being
equipped with a He-flow cryostat operating between 4 K and 600 K.
Using different light sources, different beam splitters, and
different detectors, we were able to cover the frequency range from
100 cm$^{-1}$ to 28000 cm$^{-1}$. For the analysis of the
reflectivity we derived the complex dielectric constants or the
complex conductivity by means of Kramers-Kronig transformation with
a smooth power-law extrapolation towards high wave numbers. A
constant low-frequency extrapolation has been used for the
insulating compounds at low temperatures ($T <$ 130 K) and an
additional Drude-like conductivity resulting in a Hagen-Rubens
low-frequency extrapolation has been applied at elevated
temperatures, where conductivity plays an essential role. Some
optical results in the far-infrared regime wüstite have been
published previously \cite{Kant2009}. Note that in \cite{Kant2009}
the same crystal as investigated in this work is indexed as
Fe$_{0.92}$O. As detailed in section \ref{sec:DE} the stoichiometry
rather is Fe$_{0.93}$O.

\section{Experimental results and discussion}
\label{sec:2}

\subsection{Structural and electrical characterization}
\label{sec:DE}

In the x-ray diffraction measurements at room temperature, only
peaks of the proper fcc structure were observed (space group
Fm\={3}m) and no impurity peaks above background could be detected.
From a detailed Rietveld analysis we determined the lattice spacing
\textit{a} = 4.2978 Å. In addition, we refined the site occupation
of the iron ions and found a value of 0.91. From a compilation of
literature data of lattice constants of FeO with different iron
deficiency, McCammon and Liu \cite{McCammon1984}  proposed a linear
relation between the lattice spacing and the iron deficiency
$\delta$, \textit{a}  =  4.334 -  0.478 $\delta$. In our case this
relation provides $\delta$ = 0.0757 or (1 - $\delta$) = 0.924. In a
similar manner, McCammon \cite{McCammon1992} proposed a relationship
between antiferromagnetic ordering temperature and Fe concentration.
As will be documented later, the best estimate of the magnetic
ordering temperature of the compound under consideration is
$T_{\mathrm{N}}$ = 195(2) K, which corresponds to a composition (1 -
$\delta$) = 0.94. Using this information we finally estimate the
stoichiometry of our sample to be (1 - $\delta$)  = 0.93(2) and the
wüstite sample investigated in this work is referenced as
Fe$_{0.93}$O. Assuming ideal oxygen stoichiometry, this composition suggests that charge compensation
is reached for Fe$^{2+}_{0.79}$Fe$^{3+}_{0.14}$O.

\begin{figure}[t]
\resizebox{1.0\columnwidth}{!}{%
  \includegraphics{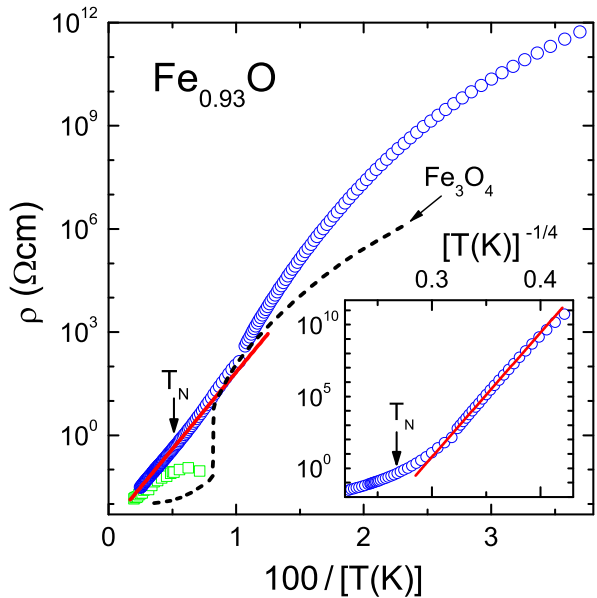}
}
\caption{Logarithm of the electrical dc resistance vs. inverse temperature of Fe$_{0.93}$O from 25 K up to 400 K in an Arrhenius representation. Resistivity data as deduced from the FIR experiments are also included (squares). The purely thermally activated behavior is indicated by the solid line. The dashed line shows the resistance of magnetite (Fe$_3$O$_4$). The inset shows the dc data as logarithm of resistivity vs. $T^{-1/4}$. The line is a fit assuming the VRH model prediction $\rho \propto \mathrm{exp}(T_0/T)^{1/4}$ \cite{Mott1969}. In the main frame and in the inset the magnetic ordering temperature is indicated by an arrow. }
\label{fig:1}      
\end{figure}

Several studies have been reported
\cite{Tannhauser1962,Geiger1966,H.K.Bowen1975}, focusing on the
electrical resistance of wüstite, specifically pointing towards the
close relationship of the temperature dependent conductivity in FeO
as compared to Fe$_3$O$_4$, which reveals a metal-to-insulator
transition close to 124 K \cite{Tannhauser1962,H.K.Bowen1975}. Bowen
et al. \cite{H.K.Bowen1975} concluded that the electrical transport
is thermally activated for temperatures above 120 K and reveals
variable-range hopping proportional to $T^{-1/4}$ at lower
temperatures. Fig. \ref{fig:1} shows the results as obtained from
the present experiments (circles). Here we have plotted the
logarithm of the resistance vs. the inverse temperature in an
Arrhenius-like plot. The measurements at high temperatures ($T
>$~100~K) have been performed in standard four-point technique. At
lower temperatures a two-point configuration has been used. As can
be seen in Fig. \ref{fig:1}, both data sets are in good agreement,
but certainly do not match perfectly. This can be ascribed to the
fact, that in four-point configuration the sample resistance becomes
of the order of the input impedance of the device, while in
two-point configuration contact resistances may play a role at
higher temperatures. The data from 130~K to 400~K can nicely be
described by purely thermally activated behavior. Using $\rho =
\rho_0 \mathrm{exp} (E_B/T)$  (solid red line in Fig. \ref{fig:1})
we find a prefactor  $\rho_0$ = 2.3 m$\mathrm{\Omega}$cm and an energy
barrier of 1030 K corresponding to 89 meV. These findings are in good agreement with the results of Bowen et al.
\cite{H.K.Bowen1975}, who investigated a wüstite sample with an iron
concentration of 0.91. The purely thermally activated behavior
extends from high temperatures deep into the magnetically ordered
state until approximately 130 K. In the inset of Fig. \ref{fig:1} we
show the same data set as logarithm of resistance vs. $T^{-1/4}$.
This should lead to linear behavior for variable-range hopping (VRH)
of localized charge carriers near the Fermi level in a disordered
semiconductor as first predicted by Mott \cite{Mott1969}. In a
limited temperature range of 40 K $< T <$ 90 K, but over
approximately six decades of conductivity, we find reasonable
agreement of the experimental results with the VRH model.

At this stage we would like to make some comments on these
interesting results. First of all, the transition from purely
thermally activated to VRH behavior has to be expected in disordered
semiconductors \cite{Mott1978}. In systems revealing this behavior
the Fermi energy lies within the region of localized states,
extending from the band edge to the mobility edge. At high
temperatures charge carriers can be excited across the mobility edge
and follow an Arrhenius-type of temperature behavior, finally
asymptotically  reaching  the so called "Mott minimum metallic
conductivity", which is given by 610/$a$ ($\mathrm{\Omega}$cm)$^{-1}$ if the
lattice constant $a$ is given in Å \cite{Mott1972}. Taking
the lattice constant of wüstite given above, leads to 142
($\mathrm{\Omega}$cm)$^{-1}$, a value in the order of magnitude as observed
in Fig. \ref{fig:1} with $\sigma_0 = 1/\rho_0$ = 435
($\mathrm{\Omega}$cm)$^{-1}$.  At low temperatures the electrons can only hop
between different localized states close to the Fermi energy and a
$T^{1/4}$ behavior can be expected. In wüstite due to cation
vacancies an acceptor band is believed to form just above the
valence band. To take into account the observation of VRH
conductivity, it seems straightforward to assume that due to disorder
the high-energy states in the valence band are localized and that a
mobility edge forms close to the upper band edge. Hence, free charge
carriers can only be expected when holes are created below this
mobility edge. Earlier, Bowen et al. \cite{H.K.Bowen1975} have assumed that the energy
gap as determined by the resistivity experiments corresponds to
transitions from the top of the valence band to the acceptor energy. We propose that the activation energy as
determined from the dc resistivity documented in Fig. \ref{fig:1} is
a measure of the distance from the mobility edge to the acceptor
level.

\begin{figure}[t]
\resizebox{1.0\columnwidth}{!}{%
  \includegraphics{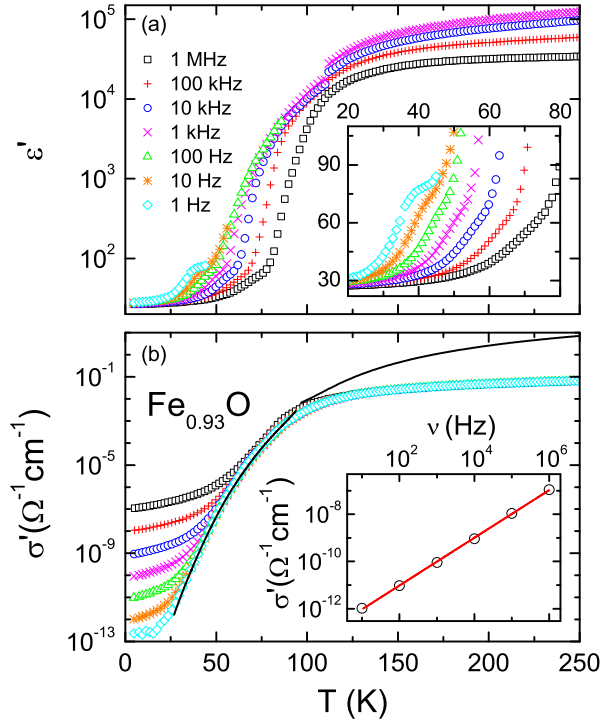}
}
\caption{Temperature dependence of the real part of the dielectric constant (a) and of the real part of the conductivity (b) for frequencies between 1 Hz and 1 MHz as function of temperature. The line in (b) represents the dc conductivity as determined from the resistivity data shown in Fig. \ref{fig:1}. The inset in (a) provides an enlarged view of the dielectric constant below 50 K. The inset in (b) is a double-logarithmic representation of the conductivity as function of frequency as observed at 5 K. The line represents a power-law behavior with an exponent $s \approx$ 1.0.  }
\label{fig:DE}       
\end{figure}

It is worthwhile to compare the temperature dependent
resistivity of wüstite with that of magnetite, Fe$_3$O$_4$. The
latter compound exhibits a metal-to-insulator transition due to
charge-ordering phenomena close to 130 K. In Fig. \ref{fig:1} we
have included the resistivity as determined in magnetite in a
sample that has been characterized in detail in Ref.
\cite{Schrettle2010}. Magnetite is characterized by Fe$^{2+}$ in
tetrahedral environment, but in addition by a distribution of
Fe$^{2+}$ and Fe$^{3+}$ in octahedral environment. The resulting
resistance is shown as dashed line in Fig. \ref{fig:1}.
Interestingly, at high temperatures in both compounds, in
"metallic" magnetite as well as in "semiconducting" wüstite, the
resistance is of the order of 10 m$\mathrm{\Omega}$cm, but with
a much weaker temperature dependence in the former case. At the
insulating side of the metal-to-insulator transition both
resistivities again become equal, close to
100~$\mathrm{\Omega}$cm and show a similar further non-Arrhenius
increase towards low temperatures.

Fig. \ref{fig:DE}a shows the real  part $\varepsilon'$ of the
complex dielectric constant of Fe$_{0.93}$O as a function of
temperature for various frequencies between 1 Hz and 1 MHz. Close to
room temperature, $\varepsilon'(T)$ is of the order of 10$^5$. Such
high values of the dielectric constant are often generated by
electrode polarization due to formation of a diode at the interface
of contact and sample \cite{Lunkenheimer2002}. Hence, the observed
high values of $\varepsilon'$ are of extrinsic origin. For all
frequencies measured, the dielectric constant decreases to a weakly
indicated lower plateau with values of $\varepsilon' \approx$ 70
(see inset of Fig. \ref{fig:DE}a). This steplike decrease is
strongly frequency dependent and can be ascribed to a non-intrinsic,
so-called Maxwell-Wagner relaxation
\cite{Lunkenheimer2002,Maxwell1873,Wagner1913}. Upon further
cooling, $\varepsilon'(T)$ decreases again to a value of
approximately 27. This second decrease also shows a significant
frequency dependence and is shifted towards lower temperatures with
decreasing frequency. We ascribe it to an intrinsic dielectric
relaxation. In non-stoichiometric wüstite with significant iron
deficiency the appearance of Fe$^{3+}$ or holes at the oxygen sites
can be expected. These entities could be responsible for the
creation of polar defects. Again we would like to stress the close
relationship with magnetite. In magnetite in a comparable
temperature range similar, but certainly much stronger relaxation
phenomena have been detected which establish a short-range polar
state at low temperatures \cite{Schrettle2010}. In wüstite we have
checked the field dependent polarization for temperatures below 50 K
and found strictly linear field dependence and
no indications of polar hysteresis effects for all temperatures. 
It is worthwhile to
mention that similar dispersion effects have been detected in
different TM compounds and interpreted as evidence for
bound polarons \cite{Lunkenheimer1992,Pimenov1994} or to be due to
orbital fluctuations \cite{Fichtl2005}. Both explanations seem to be
possible in wüstite, too. However, a detailed analysis of these
effects is severely hampered by the strong Maxwell-Wagner relaxation
which dominates the dielectric spectra above 30 K.

The real part of the conductivity, which is proportional to the
dielectric loss, is shown in Fig. \ref{fig:DE}b. At temperatures
below 50~K the conductivity reveals strong dispersion, which
indicates frequency dependent conductivity due to hopping processes
in agreement with the dc results presented in Fig. \ref{fig:1}. In
the lower inset we show the frequency dependence of the ac
conductivity at 5~K and find a dependence of the form $\sigma'
\propto \nu^s$ with $s$ close to 1.0. A frequency dependent hopping
conductivity is observed in many disordered semiconductors and the
frequency exponent in many cases approximately approaches values close to 1 
\cite{Elliott1987}. This observation might be indicative for small
polaron hopping, as it has been stated by van Staveren
\cite{vanStaveren1991}, that in contradiction to Ref.
\cite{Elliott1987}, it is the small polaron hopping that exhibits a
frequency exponent close to 1 at low temperatures. At high
temperatures the conductivity reaches a plateau and is dominated by
electrode polarization, which determines the two-point resistance in
a variety of materials as decribed in detail, e.g. in
\cite{Lunkenheimer2002,Emmert2011}. Between 25 K and 100~K the ac
conductivity as determined at 1~Hz nicely follows the dc
conductivity, which is indicated as solid line.

\subsection{Magnetic susceptibility}

The magnetic susceptibility of wüstite has been investigated by Bizette and
Tsai \cite{Bizette1943}, Ariya and Grossman \cite{Ariya1956}, Koch
and Fine \cite{Koch1967} and by Srinivasan and Seehra
\cite{Srinivasan1983}. The latter investigated the magnetic
susceptibility just around the magnetic ordering transition,
concluding that it is of first order. The other three groups aimed
to determine Curie-Weiss temperatures $\Theta$ and Curie constants
$C$ of the paramagnetic susceptibility. Fitting the high-temperature
susceptibility data in emu per mole using $\chi = C/(T - \Theta)$,
these groups reported vastly different results. In a sample of
unspecified composition, Bizette and Tsai \cite{Srinivasan1983}
found $T_{\mathrm{N}}$~=~198~K, $C$~=~6.24~emu~K/mol, and
$\Theta$~=~-~570~K, when analyzing the susceptibility up to room
temperature only. From high temperature data (100~K~$< T <$~1200~K)
of Fe$_{0.93}$O, Ariya and Grossman \cite{Ariya1956} reported $C$ =
3.60 emu K/mol and $\Theta$ = - 94 K, while Koch and Fine
\cite{Koch1967}, analyzing the magnetic susceptibility of a sample
with (1 - $\delta$)  =  0.93 between the ordering temperature and
400 K reported $C$ = 3.56 emu K/mol and $\Theta$ = - 136 K. The
discrepancies of these results may be explained by the different
temperature regimes that have been used for the data analysis, but
also by the slightly different stoichiometries of the samples and
the possible presence of magnetic impurity phases.

\begin{figure}[tbh]
\resizebox{1.0\columnwidth}{!}{
  \includegraphics{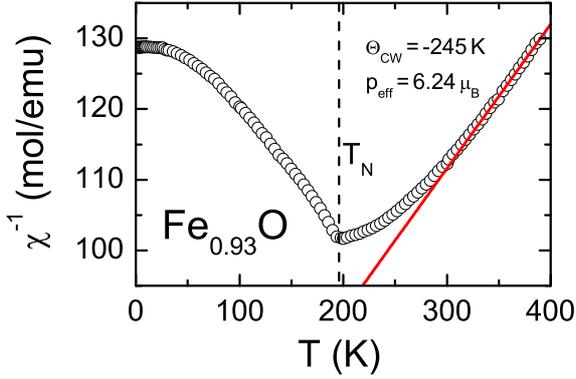}
}
\caption{Inverse magnetic susceptibility of Fe$_{0.93}$O vs. temperature as measured in an external magnetic field of $\mu_0H$~=~0.1~T \cite{Kant2009}. The solid line represents a Curie-Weiss fit between 330 K and 400 K leading to a Curie-Weiss temperature of -245 K and a paramagnetic moment of 6.49 $\mu_{\mathrm{B}}$. The magnetic ordering temperature is indicated by the dashed vertical line. }
\label{fig:2}      
\end{figure}

Fig. \ref{fig:2} shows the inverse of the susceptibility of our
sample vs. temperature between 2 K and 400 K \cite{Kant2009}. The
best estimate of $T_{\mathrm{N}}$ of this set of data is 196 K,
where a clear minimum in $1/\chi$(T) shows up. No traces of the
structural phase transition of magnetite at 124 K can be detected
and, even more importantly, any spontaneous magnetization is absent,
even at the lowest temperatures. From this observation we conclude
that our sample contains no residual $\alpha$-Fe or Fe$_3$O$_4$
clusters. Analyzing the susceptibility between 320 K and 400 K
results in a Curie constant $C$ = 4.90 emu K/mol and a Curie-Weiss
temperature $\Theta$ = - 245 K. Comparing our results with published
data, the composition of our sample is close to that investigated by
Koch and Fine \cite{Koch1967} and in addition the magnetic
susceptibilities have been analyzed in similar temperature ranges.
Calculating the effective paramagnetic moment per Fe ion, taking
into account the iron deficiency $\delta$, we obtain
$p_{\mathrm{eff}}$ = 6.26 $\mu_{\mathrm{B}}$ per Fe ion. This
effective moment has to be analyzed in terms of both Fe$^{2+}$ with
spin $S=2$ and Fe$^{3+}$, with spin $S=3/2$ according to
Fe$^{2+}_{0.79}$Fe$^{3+}_{0.14}$O. Setting the $g$-factor for
Fe$^{3+}$ with a half-filled shell equal to two, we obtain an
effective $g$-value of 2.69 for Fe$^{2+}$, which is larger than the
anisotropic $g$-values in the range $2.01< g <
2.3$ estimated in ionic crystals such as RbFeF$_3$ or FeF$_2$
\cite{Wang1966,Wang1971,Moriya1956}, but still smaller than $g\simeq
3.4$ in dilute systems with undistorted octahedral environment such
as Fe$^{2+}$ in MgO \cite{Low1960,Abragam1986}. In the latter case
the value is in accordance with the spin-orbit coupling constant
$\lambda$ = -100 cm$^{-1}$ of free Fe$^{2+}$ and a
crystal-field splitting of about 10000~cm$^{-1}$
\cite{Low1960,Abragam1986}.

Nevertheless, our value is clearly different than the ordered moment
as determined by neutron diffraction, which suggested a $g$-value
smaller than 2 \cite{Roth1958}. Kanamori calculated the paramagnetic
susceptibility for a fully stoichiometric FeO sample and predicted a
Curie constant $C$ = 3.15 emu K/mol and a Curie-Weiss temperature
$\Theta$ = - 332 K, again different from experimental results
\cite{Kanamori1957}. It seems that a consistent description of the
effective paramagnetic and the ordered moment can only be reached by
a detailed comparative studies of samples with well defined
stoichiometry.

\subsection{Heat capacity}

The heat capacity of ferrous oxide has been studied by a number of
groups, including Millar \cite{Millar1929}, Todd and Bonnickson
\cite{Todd1951}, Mainard et al. \cite{Mainard1968}, Gronvold et al.
\cite{Grnvold1993}, and Stolen et al. \cite{Stolen1996}. The main
observation of these experiments was a strong stoichiometry
dependence of the specific-heat anomaly at the magnetic ordering
temperature. While in Ref. \cite{Todd1951} a well defined and
relatively sharp anomaly is visible in a sample with $(1 - \delta)$
= 0.947, Gronvold et al. \cite{Grnvold1993} report relatively small
and smeared out anomalies for wüstite samples with iron
concentrations of 0.9254 and 0.9379. The concentration dependence of
the entropy deduced from the heat capacity anomaly at
$T_{\mathrm{N}}$ has been systematically investigated in Ref.
\cite{Mainard1968}.

\begin{figure}[t]
\resizebox{1.0\columnwidth}{!}{%
  \includegraphics{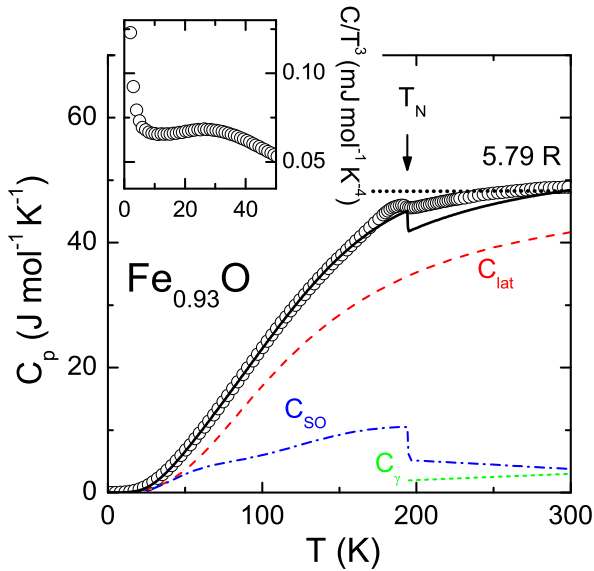}
}
\caption{Molar heat capacity of wüstite, plotted as $C_p$ vs. $T$ \cite{Kant2009}. The limitting high-temperature heat capacity value according to 5.79 R is indicated as dotted line. The antiferromagnetic ordering temperature is indicated by an arrow. The long dashed, dashed dotted and short dashed lines correspond to lattice contributions (red, long dashed), spin-orbit contributions (blue, dash dotted) and contributions due to free electrons (green, short dashed). For details see text. The inset shows $C_p/T^3$ vs. $T$ below 50 K signaling significant deviations from a pure Debye behavior. }
\label{fig:3}       
\end{figure}

The temperature of the specific heat $C_p$ of our sample, measured
at constant atmospheric pressure, is shown in Fig. \ref{fig:3}.  A
small and smeared out anomaly similarly to the data obtained in
\cite{Grnvold1993} characterizes the antiferromagnetic phase
transition, from which we determine $T_{\mathrm{N}} \approx $ 194 K.
Taking this value together with $T_{\mathrm{N}}$ = 196 K from
$\chi$(T), we arrive at an average antiferromagnetic transition
temperature of Fe$_{0.93}$O of $T_{\mathrm{N}}$ = 195(2)~K. No
further indications of phase transitions can be detected,
documenting that the sample under investigation contains no spurious
phases, specifically no traces of magnetite which would give an
anomaly close to the Verwey transition at 124~K in Fe-rich wüstite
\cite{Stolen1996}. At room temperature the heat capacity approaches
values close to 5.79 R, corresponding to the high temperature limit of the lattice specific heat only. 
Here we took into account, that the compound under consideration has an iron deficiency of 0.07 and hence is not a perfect two atomic solid. 

Plotting the specific heat as $C_p/T^3$ vs. $T$ in Fig.~\ref{fig:3}
reveals a almost constant behavior for 10~K$<T<$ 30~K and indicates
a $T^3$ dependence stemming from a Debye-like phonon density of
states and gapless AFM magnon excitations. At the lowest temperatures, $T <$~10~K, significant
deviations from this $T^3$ behavior appear in $C_p/T^3$, which might
be attributed to the presence of Fe$^{3+}$ which reportedly exhibit
a crystal-field splitting, which can lead to a
Schottky-like increase towards lowest temperatures. 
Alternatively this non-Debye like increase of $C_p/T^3$ towards low temperatures could also signal glass-like behavior due to the significant iron deficiency. 

In the following we will analyze the heat capacity data by assuming
that the total heat capacity $C_{tot}$ consists of contributions
originating from vibrational, spin, orbital, and charge degrees of
freedom. To distinguish these contributions we will rely on the
low-frequency optical excitation spectrum. A similar approach has
been used to model successfully the specific heat in CoO
\cite{Kant2009}:

We will start with the $2S+1$ spin and the three-fold degeneracy of
the orbital ground state of Fe$^{2+}$ ($t_{2g}^4 e_g^2$) in the
paramagnetic regime. The low-lying electronic Fe$^{2+}$ levels can
be described by an effective total momentum
$\mathbf{J}=\mathbf{L}+\mathbf{S}$ with $\mathbf{L}$ corresponding
to an effective orbital momentum with $L=1$ and spin $S=2$.
Spin-orbit coupling $\lambda$ leads to a triply degenerate ground
state with $J=1$, the first excited state with $J=2$ at 2$\lambda$,
and the second exited state with $J=3$ at 5$\lambda$
\cite{Goodenough1968b}. We approximate the contribution of these
levels to the specific heat by
\begin{eqnarray}\label{Cso}
C_{SO}(T)&=&N\frac{\partial E}{\partial T} \quad \mbox{and}\\
E&=&\frac{1}{Z}\sum_{i=0}^2 g_i\epsilon_i e^{-\beta \epsilon{_i}},
\end{eqnarray}
with the partition function $Z=\sum_{i=0}^2 g_i e^{-\beta
\epsilon{_i}}$, the excitation energies
$\epsilon_{0,1,2}=0,23,61$~meV as observed in the FIR spectra
discussed below, degeneracies $g_{0,1,2}=3,5,7$, and $\beta\equiv
1/k_BT$. The resulting contribution is shown as a dashed line in
Fig.~\ref{fig:3}. In the next step we subtract $C_{SO}(T>T_N)$ from
the experimental specific heat for $T>T_N$ and model the residual
specific heat in the paramagnetic regime by a sum of one isotropic
Debye, one isotropic Einstein term and a linear contribution $\gamma
T$ justified by the observation of a Drude-like contribution in the
paramagnetic state. Constraining the Debye temperature to the
vicinity of $\theta_D$~=~385~K as derived from the low-temperature
data and the Einstein temperature to vary around the experimental
eigenfrequency of the transverse polar phonon, we find a
satisfactory agreement using the parameters $\theta_D$~=~385~K,
$\theta_E$~=~479~K, and $\gamma$ = 10~mJ/molK. We further subtract
the linear contribution in the paramagnetic regime (dash-dotted
line) and the lattice contribution $C_{lat}$ (dotted line) in the
entire temperature range assuming that changes below the magnetic
transition are negligible. As a result we can write the total
specific heat as
\begin{eqnarray}
C_{tot}&=&C_{lat}(T) + \gamma T(T>T_N)  \\\nonumber
&+&C_{SO}(T>T_N)+ C_{SO}(T<T_N). 
\end{eqnarray}
Alone, the contribution $C_{SO}(T<T_N)$ has not yet been determined.
In the magnetically ordered state the degeneracy of the Fe$^{2+}$
levels will be lifted due the exchange coupling. These splittings
were not resolved in the optical experiment, but we can simulate the
residual data $C_{SO}(T<T_N)$ satisfactorily by assuming
that only the lowest-lying level is split and the higher-lying
states are only shifted in energy. The resulting $C_{SO}(T<T_N)$
(dashed line) is then again approximated using Eq.~\ref{Cso}, the
anticipated excitation energies $\epsilon_i=0,15,25,60,72$~meV for
and corresponding degeneracies $g_i=1,1,1,5,7$ for $i={0,1,2,3,4}$.
Clearly, this procedure is not unambiguous, but summing up all
modeled contributions nicely describes the experimental data of the heat capacity of wüstite and
justifies our approach to a certain extent.

\subsection{Optical properties}

Infrared reflectivity experiments on ferrous oxide with the main
focus on phonon properties and conductivity contributions of free
charge carriers at elevated temperatures have been performed by
Bowen et al. \cite{H.K.Bowen1975}, Prevot et al. \cite{Prevot1977}
and by Henning and Mutschke \cite{Henning1997}. Recently, Seagle et
al. \cite{Seagle2009} have investigated the phonon properties and
conductivity contributions of Fe$_{0.91}$O as function of
hydrostatic pressure. The absorption at higher frequencies close to
the fundamental absorption edge has only been published in Ref.
\cite{H.K.Bowen1975}.

\begin{figure}
\resizebox{1.0\columnwidth}{!}{%
  \includegraphics{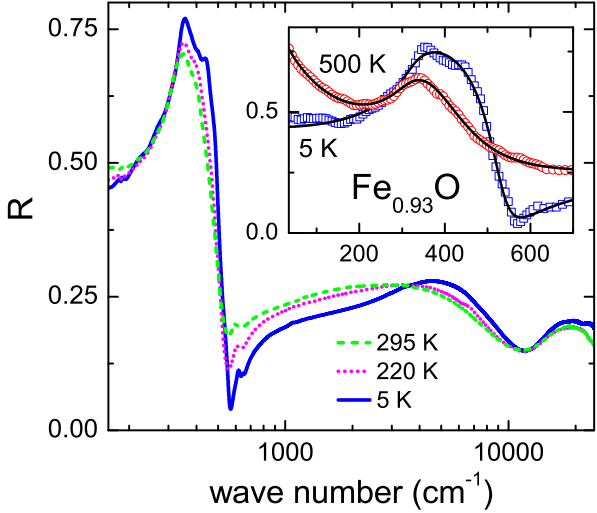}
}
\caption{Reflectivity of wüstite vs. wave number in a semi-logarithmic representation for 5 K (solid line), 220 K (dotted line) and 295 K (dashed line). The inset shows the low-frequency optical response at 5 K \cite{Kant2009} (squares) and 500 K (circles) on a linear scale. The lines represent results of fits as explained in the text. }
\label{fig:5}       
\end{figure}

In Fig.~\ref{fig:5} we have plotted the reflectivity $R$ as function
of wave number from approximately 100 cm$^{-1}$ to 24.000 cm$^{-1}$
in a semilogarithmic representation. The upper wave-number limit
corresponds to approximately 3 eV in energy. In the main frame the
reflectivity is shown at three temperatures, namely well in the
magnetically ordered state at 5 K, just above $T_{\mathrm{N}}$ at
220 K and at room temperature. From 200 cm$^{-1}$ to 700 cm$^{-1}$,
the typical phonon response of a cubic two-atomic ionic crystal
dominates the reflectivity spectra.  Electronic excitations appear
already well below 10.000 cm$^{-1}$ and approximately at 20.000
cm$^{-1}$.  The inset of Fig. \ref{fig:5} shows the low frequency
response at 5 K and at 500 K. At elevated temperatures, the strong
increase of the reflectivity towards low frequencies and the
decrease of the dipolar strength of the phonon response due to
screening effects clearly signal the existence of an increasing
number of free charge carriers.

\subsubsection{Phonon properties and Drude-like behavior}

We have converted the reflectivity into the complex dielectric
constant $\varepsilon(\omega) = \varepsilon' - i\varepsilon''$,
where $\varepsilon'$ and $\varepsilon''$ correspond to real and
imaginary part of the dielectric constant, respectively.
$\varepsilon(\omega)$ has then been analyzed using the model
dielectric function

\begin{equation}
\varepsilon(\omega) = \varepsilon_{\infty} + \varepsilon_{\infty}\Omega^{2}_{p}/(\omega^2 - i\omega/\tau) + \omega^2\Delta\varepsilon/(\omega_T^2 - \omega^2 -i\gamma\omega)
\label{eq:1}
\end{equation}

\noindent Here $\varepsilon_{\infty}$ is the electronic dielectric
constant, as defined at the upper limit of the phonon and at the
lower limit of the electronic contributions. The second term of Eq.
(\ref{eq:1}) corresponds to a Drude-like contribution due to mobile
charge carriers, with an effective plasma frequency $\Omega_p$, and
a life time $\tau$, corresponding to an inverse electronic
scattering rate. The electronic plasma frequency is defined by

\begin{equation}
\Omega_p^2 = Ne^2/(\varepsilon_0\varepsilon_{\infty}m_e)
\label{eq:2}
\end{equation}

\noindent and is directly related to the dc conductivity via

\begin{equation}
\Omega_p^2 = \sigma_{dc}/(\varepsilon_0\varepsilon_{\infty}\tau)
\label{eq:3}
\end{equation}

\noindent Here $N$ designates the charge-carriers density and $\varepsilon_0$ is the dielectric permittivity of free space. The third term in Eq. (\ref{eq:1}) describes the contributions of a normal mode to the dielectric constant according to a Lorentz oscillator. At temperatures $T < T_{\mathrm{N}}$, where the optic mode splits into two branches, the sum of two Lorentz oscillators has to be used. In Eq. (\ref{eq:1}) $\Delta\varepsilon$ designs the dielectric strength of the mode,  $\omega_T$ the transverse optical eigenfrequency  and $\gamma$ the damping of the appropriate phonon mode. Using this three-parameter fit for the phonon modes implies that the damping is frequency independent and the same for transverse (T) and longitudinal (L) modes. The longitudinal optical eigenfrequency can then be calculated via the Lyddane-Sachs-Teller relation

\begin{equation}
\varepsilon_{s}/\varepsilon_{\infty} = \left(\frac{\omega_L}{\omega_T}\right)^2
\label{eq:4}
\end{equation}

\noindent where the static dielectric constant $\varepsilon_{s}$ is given by $\varepsilon_{s} = \varepsilon_{\infty} + \Delta\varepsilon$.

For the analysis of the phonon properties and of the dynamic
conductivity one can proceed in two ways. Here we directly analyze
the reflectivity $R$ above $T_N$, which is given by

\begin{equation}
R(\omega) = \left|\frac{\sqrt{\varepsilon(\omega)} - 1}{\sqrt{\varepsilon(\omega)} + 1}\right|^2
\label{eq:5}
\end{equation}

\noindent with the complex dielectric constant as defined in Eq.
(\ref{eq:1}). The results of these fits, using eigenfrequency,
damping and dielectric strength as free parameters, are shown as
solid lines in the inset of Fig.~\ref{fig:5}. Here it should be noted that the fits in the inset of Fig. 5 have been derived by assuming one Lorentz oscillator only. This means that for this fit we have ignored the phonon splitting at 5 K. The phonon splitting in the antiferromagnetic state has been determined separately by analyzing the dielectric loss as discussed in the following. 

\begin{figure}
\resizebox{1.0\columnwidth}{!}{%
  \includegraphics{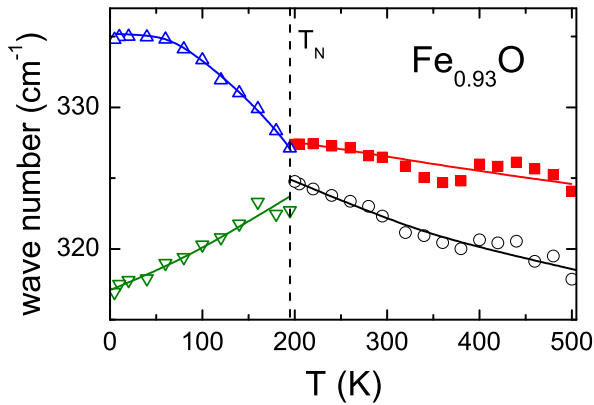}
}
\caption{Temperature dependence of the eigenfrequencies of the transverse optical mode in wüstite. The full red squares have been determined analyzing the reflectivity directly. Circles and triangles were determined from the peak maxima of the dielectric loss \cite{Kant2009}. The lines are guides to the eyes.}
\label{fig:6}       
\end{figure}

\begin{figure}
\resizebox{1.0\columnwidth}{!}{%
  \includegraphics{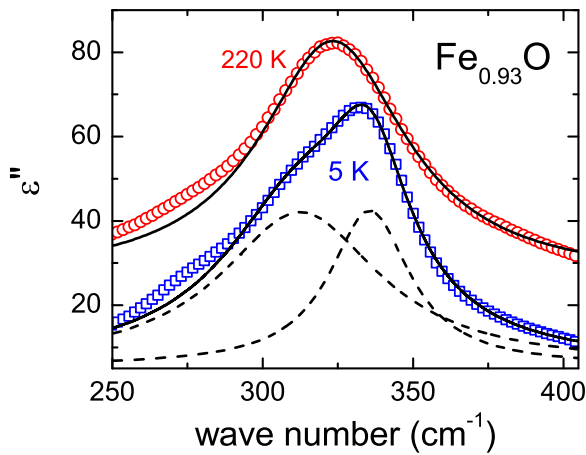}
}
\caption{Dielectric loss of wüstite vs. temperature in the AFM phase at 5 K (squares) and in the paramagnetic state at 220 K (circles) \cite{Kant2009}. The fits (solid lines) have been performed with two, respective one Lorentz oscillator. The contributions of the two Lorentz oscillators, which have been used at 5 K, are separately indicated as dashed lines. For clarity reasons, the 220 K curve has been shifted upwards by 20. }
\label{fig:7}      
\end{figure}

The resulting temperature dependence of the transverse phonon eigenfrequencies are
shown in Fig. \ref{fig:6} as (red) squares, which only show a weakly
temperature dependent phonon mode frequency of about 326 cm$^{-1}$.
This data is compared to the one obtained by the second approach,
namely the evaluation of the dielectric loss obtained after
Kramers-Kronig transformation. The results of this analysis have
been reported previously \cite{Kant2009} and, therefore, will only
shortly be summarized for completeness: An example of the dielectric
loss in wüstite is shown in Fig. \ref{fig:7}, presenting
$\varepsilon''(\omega)$ vs. wave number between 250 cm$^{-1}$ and
400 cm$^{-1}$ at 5 K, in the AFM state and at 220 K, just above the
magnetic ordering temperature in the paramagnetic phase. While the
dielectric loss at 220 K clearly shows a symmetric peak, a strong
asymmetry evolves below the onset of magnetic order. The peak at 220
K can nicely be fitted with a single Lorentz oscillator, while two
oscillators, indicated by dashed lines, are necessary to describe
the low-temperature data. Deviations from the Lorentz fits appear
close to 270 cm$^{-1}$. These deviations are almost temperature
independent and we have no explanation of their origin so far.

The splitting of the transverse optical phonon itself has been
attributed to the effects of an mostly exchange-induced coupling mechanism
\cite{Yamashita2000,Tchernyshyov2002,Massidda1999,Luo2007,Kant2011}.
The eigenfrequencies which have been determined from these fits to
the dielectric loss of wüstite are also plotted in Fig.~\ref{fig:6}.
In the paramagnetic phase the eigenfrequencies reveal a similar
temperature dependence as those determined from the fits to the
reflectivity, but are shifted by approximately 5 cm$^{-1}$ to lower
energies. This deviation gives a rough estimate of the precision of
the two different methods to derive the eigenfrequencies and
probably results from the assumption of a frequency independent
damping and from the fact that the dc conductivity has not been
equally taken into account by the two different fitting scenarios.
Below $T_{\mathrm{N}}$, in Fig. \ref{fig:6} the splitting of the
modes is clearly visible and increases on decreasing temperature. At
5 K the splitting is of the order of 15~cm$^{-1}$ and significantly
smaller than the phonon splitting observed in isostructural and AFM
MnO, where it amounts almost 25 cm$^{-1}$ \cite{Rudolf2008}.

In the following we discuss the obtained fit parameters, namely the
eigenfrequencies of transverse and longitudinal optical modes, the
damping, which is assumed to be identical for both modes, and the
static as well as the electronic or infinite dielectric constant
(table \ref{tab:1}).

\begin{table}
\caption{Phonon eigenfrequencies of transverse and longitudinal optical phonons, damping and dielectric constants at room temperature.}
\label{tab:1}       
\begin{tabular}{lllll}
\hline\noalign{\smallskip}
$\omega_{TO}$ & $\omega_{LO}$ & $\gamma$ & $\varepsilon_s$ & $\varepsilon_{\infty}$  \\
\noalign{\smallskip}\hline\noalign{\smallskip}
322.3 cm$^{-1}$ & 466.0 cm$^{-1}$ & 76.1 cm$^{-1}$ & 22.6 & 10.8 \\

\noalign{\smallskip}\hline
\end{tabular}
\end{table}

The eigenfrequencies are only partly in agreement with the results
obtained by inelastic neutron scattering \cite{Kugel1977}. At room
temperature, the zone-center optical eigenfrequencies were found to
be $\omega_T$ = 320 cm$^{-1}$ and $\omega_L$ = 526 cm$^{-1}$. While
the former certainly agrees within experimental uncertainties with
the present value of 322 cm$^{-1}$, the disagreement of the
longitudinal eigenfrequencies is much larger. In the neutron work
unpublished IR results are cited with $\omega_T$ = 290 cm$^{-1}$ and
$\omega_L$ = 535 cm$^{-1}$. From further infrared reflectivity
measurements the values of $\omega_T$ = 333.7 cm$^{-1}$ and
$\omega_L$ = 493.5 cm$^{-1}$ have been determined \cite{Prevot1977}.
In the latter work dielectric constants at zero and infinite
frequencies are published with values of $\varepsilon_s$ = 32.8 and
$\varepsilon_{\infty}$ = 9.63, respectively. The latter value is of
the order of magnitude as determined here, but nevertheless seems
quite large for transition metal monoxides. In MnO, CoO and NiO the
electronic dielectric constant, $\varepsilon_{\infty} \approx$ 5, is
by a factor of two smaller than in FeO. We think that the large
value of $\varepsilon_{\infty}$ in wüstite results from the
crystal-field excitation which significantly contributes to the
reflectivity at 1000 cm$^{-1}$. In addition, at room temperature,
and even more at elevated temperatures, conductivity contributions
hamper an exact determination of the electronic dielectric constant.
This uncertainty in the determination of $\varepsilon_{\infty}$
could also provide a possible explanation for the enormous scatter
of reported values of the longitudinal eigenfrequency. In the
dielectric measurements of this work (section \ref{sec:DE}) we have
determined a high-frequency value of the dielectric constant of 27,
which is not too far from $\varepsilon_{s} = 22.6$ determined in the
FIR experiments. If we take $\varepsilon_s$ as determined from the
dielectric measurement and $\varepsilon_{\infty}$ as well as
$\omega_T$ as determined from the infrared experiments and listed in
Table \ref{tab:1}, we determine an longitudinal optical
eigenfrequency via Eq. (\ref{eq:4}), $\omega_L$ = 509.6 cm$^{-1}$.

From the fits using Eq. (\ref{eq:1}) we also have deduced the
relevant parameters of the free-carrier contribution. This was
possible only for temperatures $T >$ 150 K. Below 150 K, in the
hopping regime (section \ref{sec:DE}), no conductivity contributions
can be determined from the FIR experiments. For low temperatures
wüstite behaves like an insulator. A closer inspection of the main
frame of Fig. \ref{fig:5} also documents that electronic transitions
dominate the optical response even well below 1 eV ($\approx$ 8000
cm$^{-1}$) and, hence, hamper the determination of the electronic
relaxation time $\tau$. From the fits as indicated in the inset of
Fig. \ref{fig:5}, only the dc conductivity can be determined,
proportional to the product $\Omega^2\tau$ (see Eq. (\ref{eq:3})).
The inverse dc conductivity as determined from these fits is plotted
in Fig. \ref{fig:1}, in addition to the resistivity results. While
the data are in reasonable agreement for high temperatures,
significant deviations appear on cooling. From Fig. \ref{fig:1} it
is clear that the dc conductivity, as determined from the
reflectivity measurements at high frequencies, follows a
significantly different Arrhenius behavior compared to $\sigma_{dc}$
determined from the low-frequency experiments. On further cooling
additional deviations appear, which probably stem from the fact that
for low temperatures the dc conductivity was too small and the
reflectivity measurements did not cover sufficiently low frequencies
to correctly determine the conductivity contributions. A fit to the
high-frequency optical results of the dc resistivity as shown in
Fig. \ref{fig:1} yields an energy barrier of 47.5 meV, almost a
factor of 2 lower than the gap determined from the dc resistivity
experiments.

\subsubsection{Electronic transitions}

\paragraph{Splitting of the ground state by spin-orbit coupling}
\label{sec:SpinOrbitCoupling}

In addition to the phononic and Drude contribution, the FIR spectra
bear fingerprints of local Fe$^{2+}$ transitions, which have been
used already above to model the specific heat. In a $d^6$ system in
an octahedral crystal field with a $^{5}T_2$ ground state ($S$ = 2;
$t_{2g}^4 e_g^2$) we expect the splitting of the ground-state
triplet into 3 levels, with the ground state at 0 and excited levels
at 2 $\lambda$ and 5 $\lambda$, so that the overall splitting is
given by 5 $\lambda$, with $\lambda$ being the spin-orbit coupling
constant \cite{Goodenough1968b}. To the best of our knowledge no
optical information on the low-lying Fe$^{2+}$ levels and the
spin-orbit coupling is available in literature. Alone, an optical
absorption line of dilute Fe$^{2+}$ in MgO has been reported close
to 10000 cm$^{-1}$ \cite{Low1960b}.

\begin{figure}
\resizebox{1.0\columnwidth}{!}{%
  \includegraphics{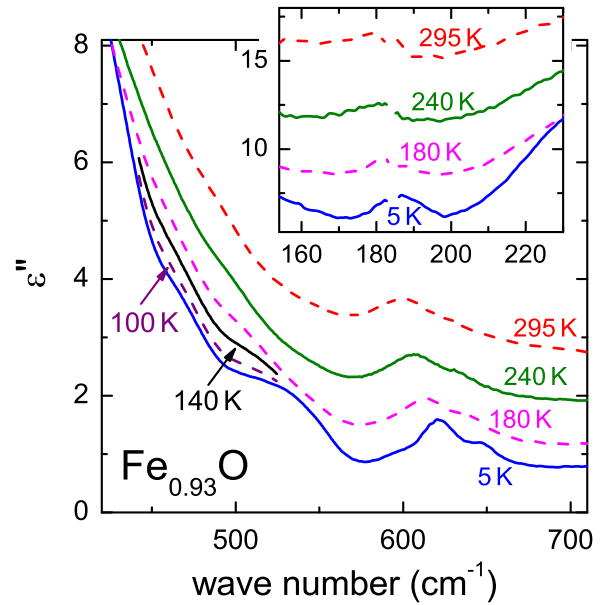}
}
\caption{Dielectric loss in wüstite between 430 cm$^{-1}$ and 700 cm$^{-1}$ for temperatures between 5 K and 295 K. The inset shows the dielectric loss for the same temperature range between 160 cm$^{-1}$ and 225 cm$^{-1}$}
\label{fig:10}       
\end{figure}

As documented in Fig. \ref{fig:10} (inset) we identify a weak
structure close to 185 cm$^{-1}$ which is almost temperature
independent and in addition we detect a tiny anomaly close to 490
cm$^{-1}$ at 240 K (main frame). This anomaly seems to split in the
magnetically ordered state and becomes smeared out at higher
temperature due to conductivity contributions to the dielectric
loss. In addition, we find an anomaly close to 600 cm$^{-1}$ at room
temperature which strongly shifts to higher frequencies on cooling.
This mode has been analyzed in detail by Seagle et al.
\cite{Seagle2009}. From the observed pressure dependence they
concluded that this mode corresponds to a localized defect
excitation. Hence, we identify the two remaining excitations close
to 185 and 490 cm$^{-1}$ as transitions between the ground state
split by spin-orbit coupling. Assuming a spin-orbit coupling
constant $\lambda$ of 95 cm$^{-1}$ we expect two transitions close
to 190 cm$^{-1}$ and 475 cm$^{-1}$ in good agreement with the
observed transitions.

\paragraph{Band gap and crystal-field excitations}
\label{sec:BandGapAndCrystalFieldExcitations}

\begin{figure}
\resizebox{1.0\columnwidth}{!}{%
  \includegraphics{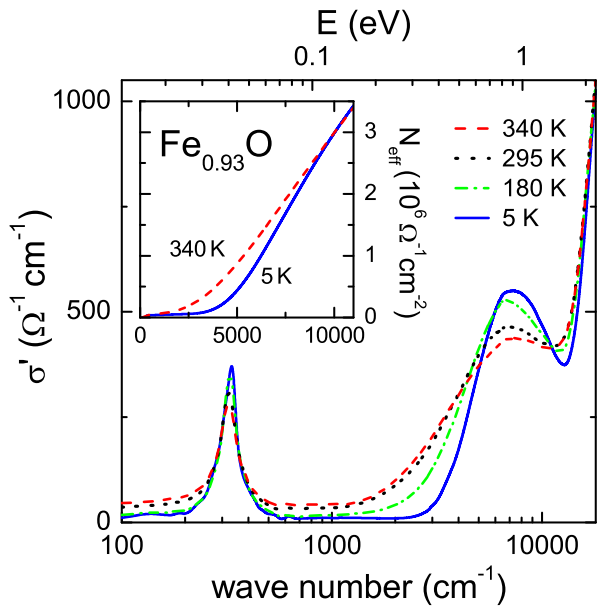}
}
\caption{Optical conductivity of wüstite vs. the logarithm of wave number at temperatures between 5 K and 340 K. The inset shows the optical weight up to frequencies just including the crystal field excitation at 5 K and 340 K.}
\label{fig:8}       
\end{figure}

The optical conductivity as derived from the  reflectivity (see Fig.
\ref{fig:5}) is documented in Fig. \ref{fig:8} for temperatures
between 5 K and 340 K. The real part of the conductivity is given by
$\sigma' = \omega\varepsilon_0\varepsilon''$. The transverse optical
phonon mode close to 320 cm$^{-1}$ is followed by a well defined
electronic transition close to 7500 cm$^{-1}$ and the onset of a
charge-transfer excitation close to 12000 cm$^{-1}$.

The electronic transition close to 7500 cm$^{-1}$ is assigned to the
spin-allowed crystal-field excitation from the $^5T_{2g}$ ground
state to the $^5E$ state which are separated by 10$Dq$ in good
agreement with the reported absorption spectra in the $A$FeX$_3$
systems, where $A$ = Cs, Rb and $X$=F, Cl, Br \cite{Putnik1976}. In
these materials a phonon-assisted mechanism has been suggested to
explain the observation of the parity-forbidden $d$-$d$ transition.
A similar mechanism might be at work in iron oxide, however, the
spectral weight of the excitation is much larger than expected for a
conventional phonon-assisted $d$-$d$ transition
\cite{Sugano1970}. Moreover, the lineshape of this excitation
in wüstite looks astonishingly similar to the excitation which has
been observed in magnetite at about 5000 cm$^{-1}$
\cite{Park1998,Gasparov2000} and analyzed in terms of a polaronic
excitation, which has been observed and discussed in many
transition-metal systems where electron-phonon coupling plays an
important role, such as e.g. manganites
\cite{Millis1996,Hartinger2004,Hartinger2006} and
cuprates \cite{Calvani2001}. Consequently, we tried to describe the
lineshape of this excitation in FeO by a small-polaron model
\cite{Emin1993}, but could not obtain a satisfactory agreement with the
data. Nevertheless, we want to recall that, similarly to the case of
magnetite, in our sample both Fe$^{2+}$ and Fe$^{3+}$ ions are
present, and the enhanced spectral weight of the crystal-field
excitation in FeO might have its origin in the presence of
mixed-valence Fe ions.

The intensity and shape of this excitation in FeO are strongly 
temperature dependent. At 5 K the conductivity up to 3000 cm$^{-1}$ is very
low and the excitation is narrow and well defined. On increasing
temperature it broadens and transfers a significant part of its
optical weight to lower frequencies, concomitantly with the increase
of the \textit{dc} conductivity contributions on increasing
temperatures. At the same time the phonon intensity becomes strongly
reduced a fact that results from the screening of the phonon modes
by free charge carriers. In the inset of Fig. \ref{fig:8} we show
the spectral weight $N_{\mathrm{eff}}$ up to wave numbers of 11000
cm$^{-1}$ at 5 K and 340 K.  The optical weight is given by the
integral of the conductivity up to a given frequency.

As expected we find exactly the same optical weight when integrating
the conductivity up to frequencies just covering the crystal field
excitations, documenting that the optical weight, which is lost by
the crystal field excitations indeed is shifted into the dc
conductivity at lower frequencies. This fact also provides
experimental evidence that a constant number of charge carriers is
responsible for the dc conductivity as well as the large spectral
weight of the crystal-field excitation. Similarly to magnetite this
relation between the crystal-field excitation and the dc
conductivity is reminiscent of polaron physics. With increasing
temperatures bound polarons become mobile. Shape and temperature
dependence of the electronic excitation depicted in Fig. \ref{fig:8}
essentially resemble observations in the colossal resistance
manganites revealing the melting of polarons into a Fermi liquid
state when passing the ferromagnetic transition from above
\cite{Millis1996}. However there is no one-to one correspondence: In
the case of the manganites, mobile polarons establish a metallic
low-temperature state, but in wüstite a paramagnetic semiconducting
state with strongly increased mobility is formed at high
temperatures.

\begin{figure}
\resizebox{1.0\columnwidth}{!}{%
  \includegraphics{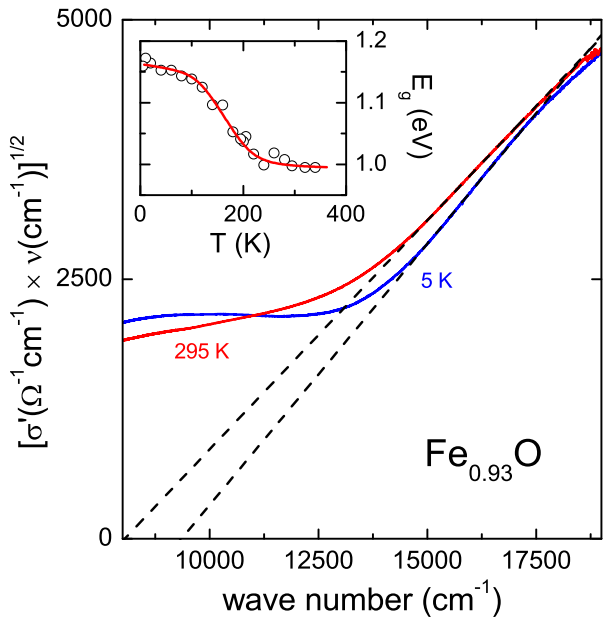}
}
\caption{Determination of gap energy for two example temperatures (solid lines) assuming an indirect allowed transition \cite{Mott1979}. The dashed lines demonstrate linear behavior, which has been extrapolated to read off the gap energies. The inset shows the resulting fundamental absorption edge $E_g$ of wüstite as function of temperature (circles). The line is a guide to the eyes.}
\label{fig:9}       
\end{figure}

The optical transition across the semiconducting band gap observed
close to 12000 cm$^{-1}$ probably corresponds to a transition from
the lower $t_{2g}$ triplet of Fe$^{2+}$ to the oxygen $p$ bands. If
this assignment is correct, this transition is a typical charge
transfer transition, as predicted to occur in the late transition
metal monoxides \cite{Zaanen1985}. From an optical absorption
experiment  Bowen et al. determined a value of 2.4 eV by evaluating
the onset of the absorption edge \cite{H.K.Bowen1975}, which is in
agreement with the theoretical estimate of 2.1~eV \cite{Lee1991}. In
our data (see Fig. \ref{fig:8}) the onset also appears at about
2~eV. We analyzed the slope of this onset and found that the best
fits can be obtained assuming an indirect allowed electronic
transition \cite{Mott1979}, which is documented in Fig. \ref{fig:9}
by plotting $(\sigma' \times \nu)^{1/2}$ vs wave number. By
extrapolating this behavior (dashed lines) we find a value of
the band gap $E_g \approx$ 1.0 eV above the onset of AFM order.
Using this extrapolation procedure we want to discuss the
temperature dependence of the band-gap as determined from our data
in the following.

\paragraph{Blue shift of the fundamental absorption edge}
\label{sec:BlueShiftOfTheFundamentalAbsorptionEdge}

In the inset of Fig. \ref{fig:9} we document the temperature
dependence of $E_g$. The gap energy is constant and approximately
1.0 eV in the paramagnetic phase and reveals a strong increase up to
1.15 eV in the AFM state. This blue shift of the band gap in the
magnetically ordered state amounts to approximately 15\%.

Shifts of absorption edges upon magnetic ordering have been studied
intensively  in the case of magnetic semiconductors. Systems
dominated by ferromagnetic exchange rather reveal strong red shifts
such as e.g. Eu chalcogenides
\cite{Busch1964,Argyle1965,Busch1966b,Suits1966,Methfessel1967,Busch1967}.
or the chromium spinels CdCr$_2$Se$_4$
\cite{Busch1966,Harbeke1966,Harbeke1970} and HgCr$_2$Se$_4$
\cite{Lehmann1969}, as well as for ZnCr$_2$Se$_4$ \cite{Busch1966},
and HgCr$_2$S$_4$, which order antiferromagnetically but posses
strong ferromagnetic exchange interactions
\cite{Harbeke1968,Lehmann1970}. An exception seems to be
CdCr$_2$S$_4$, where a blue shift has been observed
\cite{Busch1966,Harbeke1966}, but it has been argued that the
evaluated absorption feature is not the fundamental absorption edge,
but rather the low-energy flank of a crystal-field excitation
\cite{Wittekoek1969,Berger1969,Rudolf2009}. The red shifts in the
ferromagnetic semiconductors have been theoretically tackled by
Callen \cite{Callen1968} using the concept of magnetoelastic
coupling, by Baltensperger \cite{Baltensperger1970} assuming that
magnetic correlations influence the band states via exchange
coupling and by Nolting \cite{Nolting1977} calculating the influence
of the localized Heisenberg spins on the conduction band.

Blue shifts at the AFM phase transition have been reported and
analyzed for compounds related to wüstite, namely for $\alpha$-MnS
and CoO \cite{Chou1974,Terasawa1980}, and for hexagonal MnTe
\cite{FerrerRoca2000} or NaCrS$_2$ \cite{Blazey1969}. However, also
red shifts can occur as in AFM MnO \cite{Chou1974}, and the shift
seem to depend strongly on the different exchange interactions and
magneto-structural effects and distortions of the particular
compound \cite{Chou1974,Alexander1976,FerrerRoca2000}. The question
which model is appropriate to explain the significant blue shift in
passing the antiferromagnetic phase transition in wüstite remains to
be answered.

\section{Discussion and concluding remarks}

In this work we have provided a detailed study of the optical
properties of wüstite with stoichiometry Fe$_{0.93}$O. In addition
we have presented a detailed characterization of the sample with
respect to electrical, magnetic, and thermodynamic properties. From
electrical resistance experiments we find a purely thermally
activated temperature dependence governed by an energy gap of 89 meV
at elevated temperatures ($T >$~150~K). At lower temperatures
hopping conduction dominates. We conclude that charge transport in
wüstite results from charge carriers located in the valence band
below a mobility edge. Holes can be created by thermal excitations
of electrons into an acceptor level just above the Fermi energy. In
dielectric measurements we have determined a limiting high-frequency
dielectric constant of 27. Moreover, we found clear indications of a
relaxational process similar but significantly weaker when compared
to that observed in magnetite \cite{Schrettle2010}. One could
speculate on charge ordering processes between Fe$^{2+}$ and
Fe$^{3+}$ or local relaxation phenomena of polar vacancies. However,
no evidence of short or long range polar order could be detected in
wüstite. For temperatures $T >$ 80 K the dielectric constant
increases to colossal values, a fact that can be explained in terms
of Maxwell-Wagner relaxation.

Measurements of the magnetic susceptibility indicate pure
Curie-Weiss behavior, indicative of localized moments with a
Curie-Weiss temperature of $\Theta$ = - 245 K and a magnetic
ordering temperature of $T_N$ = 196 K. The temperature dependence of
the heat capacity reveals a smeared-out anomaly close to the onset
of antiferromagnetic order and could be nicely described by using
the spectroscopic data obtained from the optical experiment.

The main focus of this work was dedicated to optical properties.
Wüstite behaves like an insulator for temperatures below the Nèel
temperature and reveals increasing conductivity contributions for
higher temperatures. At 500~K a small Drude component in the
conductivity and screening effects of the phonon excitations signal
increasing metallic behavior. The observed transverse optical phonon
modes have eigenfrequencies close to 325~cm$^{-1}$. Below the AFM
ordering temperature the TO branch splits into two modes with a mode
splitting close to 15~cm$^{-1}$. From reflectivity experiments up to
3~eV we observed the $^5T_{2g}\rightarrow ^5E$ crystal-field
excitation of the Fe$^{2+}$ ions at about 7500~cm$^{-1}$ yielding a
crystal field parameter $Dq$~=~750~cm$^{-1}$. In addition we provide
experimental evidence for two further electronic Fe$^{2+}$
excitations close to 185 and 490~cm$^{-1}$ which are interpreted as
transitions between the ground-state levels split by spin-orbit
coupling. From this observation we conclude that wüstite is
characterized by a spin-orbit coupling constant $\lambda$ $\approx$
95~cm$^{-1}$.

Finally, we found a strong blue shift of the fundamental absorption
edge on cooling across the antiferromagnetic phase transition. The
blue shift amounts 0.15 eV which is an effect as large as 15\%.

Acknowledgement: This work has partly been supported by the DFG via
the Transregional Research Collaboration TRR 80 (Augsburg/Munich).

 \bibliographystyle{epjtest3}
 \bibliography{FeOBIB}

\end{document}